\documentclass[aps,pra,twocolumn,superscriptaddress,showpacs,floatfix]{revtex4}
\usepackage{amsmath,amssymb,amsfonts,bbm,graphicx,hyperref,times,color,mathptmx}
\usepackage{subfigure}
\newcommand{\ket}[1]{\vert #1 \rangle}
\newcommand{\bra}[1]{\langle #1 \vert}
\newcommand{\ave}[1]{\langle #1 \rangle}

\newcommand{\ketbra}[2]{\vert #1 \rangle \langle #2 \vert}
\def\Tr{\hbox{Tr}}

\begin{document}
\title{
Assessing the significance of fidelity as a figure of merit in quantum
state reconstruction of discrete and continuous variable systems}
\author{Antonio Mandarino}
\affiliation{Dipartimento di Fisica, Universit\`a degli Studi di
Milano, I-20133 Milano, Italy}
\author{Matteo Bina}
\affiliation{Dipartimento di Fisica, Universit\`a degli Studi di
Milano, I-20133 Milano, Italy}
\author{Carmen Porto}
\affiliation{Dipartimento di Fisica, Universit\`a degli Studi di
Milano, I-20133 Milano, Italy}
 \author{Simone Cialdi}
 \affiliation{Dipartimento di Fisica, Universit\`a degli Studi di
 Milano, I-20133 Milano, Italy}
 \affiliation{Istituto Nazionale di Fisica Nucleare, Sezione di 
 Milano, I-20133 Milan, Italy}
\author{ Stefano Olivares}
\affiliation{Dipartimento di Fisica, Universit\`a degli Studi di
Milano, I-20133 Milano, Italy}
\affiliation{Istituto Nazionale di Fisica Nucleare, 
Sezione di Milano, I-20133 Milan, Italy}
\author{ Matteo G. A. Paris}
 \affiliation{Dipartimento di Fisica, Universit\`a degli Studi di
 Milano, I-20133 Milano, Italy}
\affiliation{Istituto Nazionale di Fisica Nucleare, 
Sezione di Milano, I-20133 Milan, Italy}
\date{\today}
\begin{abstract}
We experimentally address the significance of fidelity as a figure of
merit in quantum state reconstruction of discrete (DV) and continuous
variable (CV) quantum optical systems.  In particular, we analyze the
use of fidelity in quantum homodyne tomography of CV states and
maximum-likelihood polarization tomography of DV ones, focussing
attention on nonclassicality, entanglement and quantum discord as a
function of fidelity to a target state.  Our findings show that high
values of fidelity, despite well quantifying geometrical proximity in
the Hilbert space, may be obtained for states displaying opposite
physical properties, e.g.  quantum or semiclassical features.  In
particular, we analyze in details the quantum-to-classical transition
for squeezed thermal states of a single-mode optical 
system and for Werner states of a two-photon polarization qubit system.
\end{abstract}
\maketitle
\section{Introduction}
In quantum technology, it is very common to summarize 
the results of a reconstruction technique, either full quantum 
tomography \cite{revt1,LNP649,revt2,guhne} or some partial reconstruction 
scheme \cite{jay57,buz98,oli07,zam05,kb03,olom03}, 
by the use of fidelity \cite{Uhl,fuc99}. 
Once the information about the state of a system has been extracted 
from a set of experimental data, 
the fidelity between the reconstructed state and a
given target state, is calculated.  Fidelity is bounded to the interval
$[0, 1]$. High values such as $0.9$ or $0.99$ are thus considered as a
piece of evidence in order to certify that the reconstructed and the
target states i) are very close each other in the Hilbert space, ii) 
they share nearly identical physical properties. In this framework, 
quantum resources of the prepared state are often benchmarked
with those of the target one, e.g. to assess the performances of a 
teleportation scheme \cite{ban04,cav04}. 
\par
The two statements above may appear rather intuitive, with the second 
one following from the first one. On the other hand, it has been 
suggested that the use of fidelity 
may be misleading in several situations involving either discrete 
or continuous variable systems \cite{Dod12,AgFid1,AgFid2,Benedetti}.
The main goal of the present paper is to experimentally confirm 
the first statement and, at the same time, to provide neat examples 
where the second one is clearly proved wrong.
\par
Given two quantum states described by density matrices $\hat{\rho}_1$ and
$\hat{\rho}_2$, the fidelity between them is defined as \cite{Uhl}
\begin{equation} \label{Fidelity} F(\hat{\rho}_1, \hat{\rho}_2)
=\Tr\left[\sqrt{ \sqrt{\hat{\rho}_1 } \hat{\rho}_2 \sqrt{\hat{\rho}_1
}}\right]^2\,.\end{equation} 
Fidelity is not a proper distance in the
Hilbert space. However, it can be easily linked to a distance, and
in turn to a metric over the manifold of density matrices. In fact, 
the Bures distance between two states is defined as 
$$D_B(\hat{\rho}_1,\hat{\rho}_2)=\sqrt{2[
1-\sqrt{F(\hat{\rho}_1,\hat{\rho}_2)}]}\,.$$ 
Fidelity also provides an upper and a lower bound 
to the trace distance, namely \cite{fuc99}:
$$1-\sqrt{F(\hat{\rho}_1,\hat{\rho}_2)}\leq \frac12 ||
\hat{\rho}_1-\hat{\rho}_2||_1\leq
\sqrt{1-F(\hat{\rho}_1,\hat{\rho}_2)}\,.$$
These relationships ensure that higher values of fidelity correspond to
geometrical proximity of the two states in the Hilbert space.  However,
they do not seem straightforwardly related to the physical properties of
the two states.  In turn, it has been pointed out
\cite{Dod12,AgFid1,AgFid2,Benedetti} that a pair of states that appear
very close to each other in terms of fidelity, may be very far in terms
of physical resources. Relevant examples may be found with bipartite
systems of either qubits or CV Gaussian states, where pairs of states composed
by one entangled and one separable states may have (very) high value of
fidelity one to each other.  Besides, for single-mode CV states high
values of fidelity may be achieved by pairs including one state with a
classical analogue and a genuinely quantum state of the field. 
\par
In this paper, we address the problem experimentally and analyze in
details the significance of fidelity as a figure of merit to assess the
properties of tomographycally reconstructed state.  We address both
discrete and continuous variable systems using quantum homodyne
tomography to reconstruct CV states and maximum-likelihood polarization
tomography for DV ones.  In particular, we experimentally address two
relevant examples: i) the reconstruction of squeezed thermal states of a
single-mode radiation field, analyzing in details the
quantum-to-classical transition; and ii) the reconstruction of noisy
Werner states of a two-qubit polarization system, inspecting the
amount of non-classical correlations.  Our results clearly show that
high values of fidelity, despite well quantifying geometrical closeness
between states in the Hilbert space, may be obtained for quantum states
displaying very different physical properties, e.g. quantum resources.
\par
The paper is structured as follows. Sect.~\ref{s:STS} is devoted to
continuous variables: we first describe the experimental generation of
single-mode squeezed thermal states using a seeded optical amplifier, as
well as the homodyne technique employed for tomography.  We then present
experimental results, illustrating in details the significance of the
fidelity of the reconstructed state to a target one and its
non-classicality. In Sect.~\ref{s:Werner} we illustrate the
experimental setup for generating two-qubit states of Werner type and
the method of maximum-likelihood estimation for tomography. We then
present experimental results, analysing the significance of fidelity of
the reconstructed states to the target Werner ones in assessing their
non-classical correlations, either entanglement or quantum discord.
Sect.~\ref{s:conclusions} closes the paper with some concluding remarks.
\section{Single-mode Gaussian states}\label{s:STS}
\begin{figure}[t]
\subfigure[]{\includegraphics[width=0.99\columnwidth]{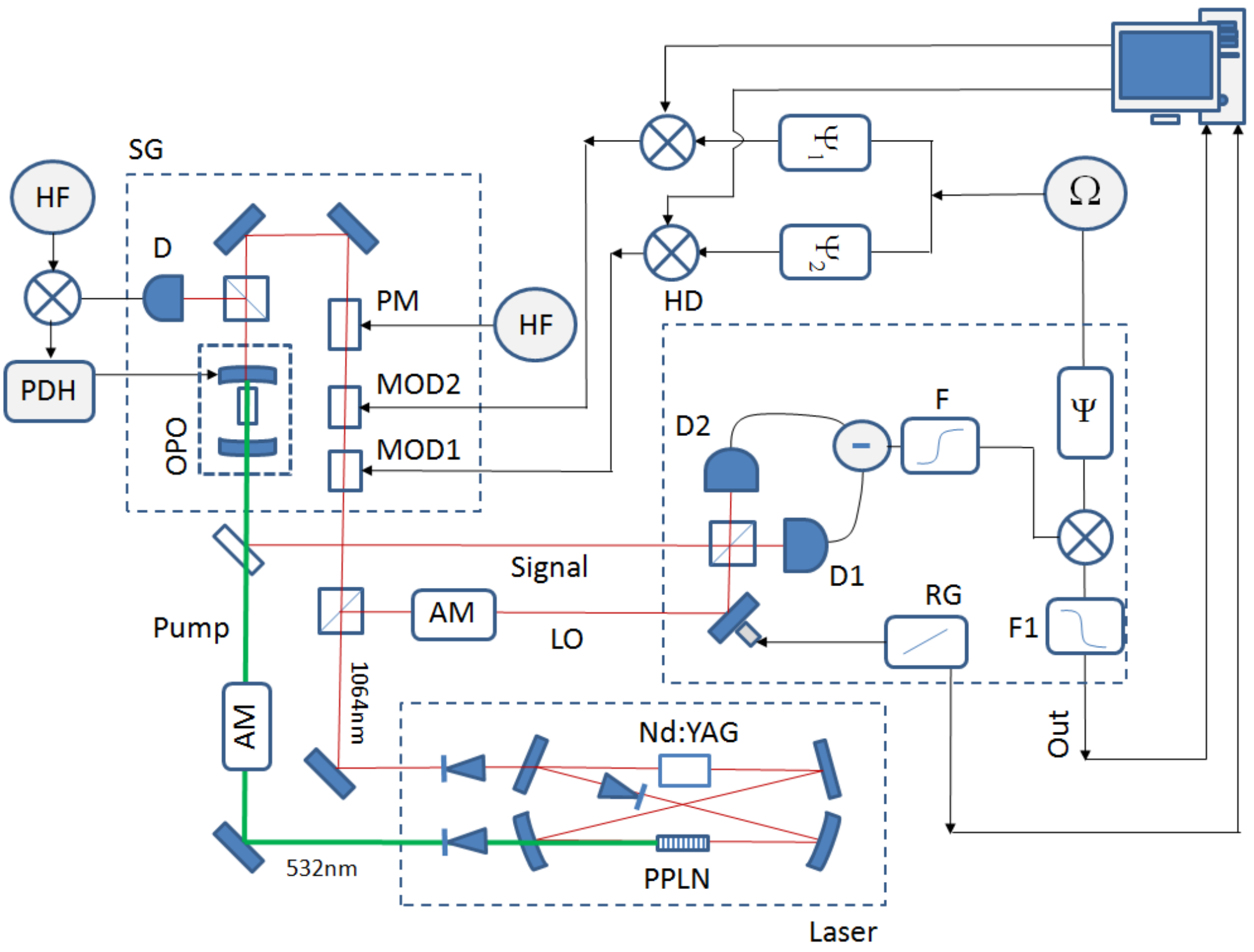}}
\subfigure[]{\includegraphics[width=0.85\columnwidth]{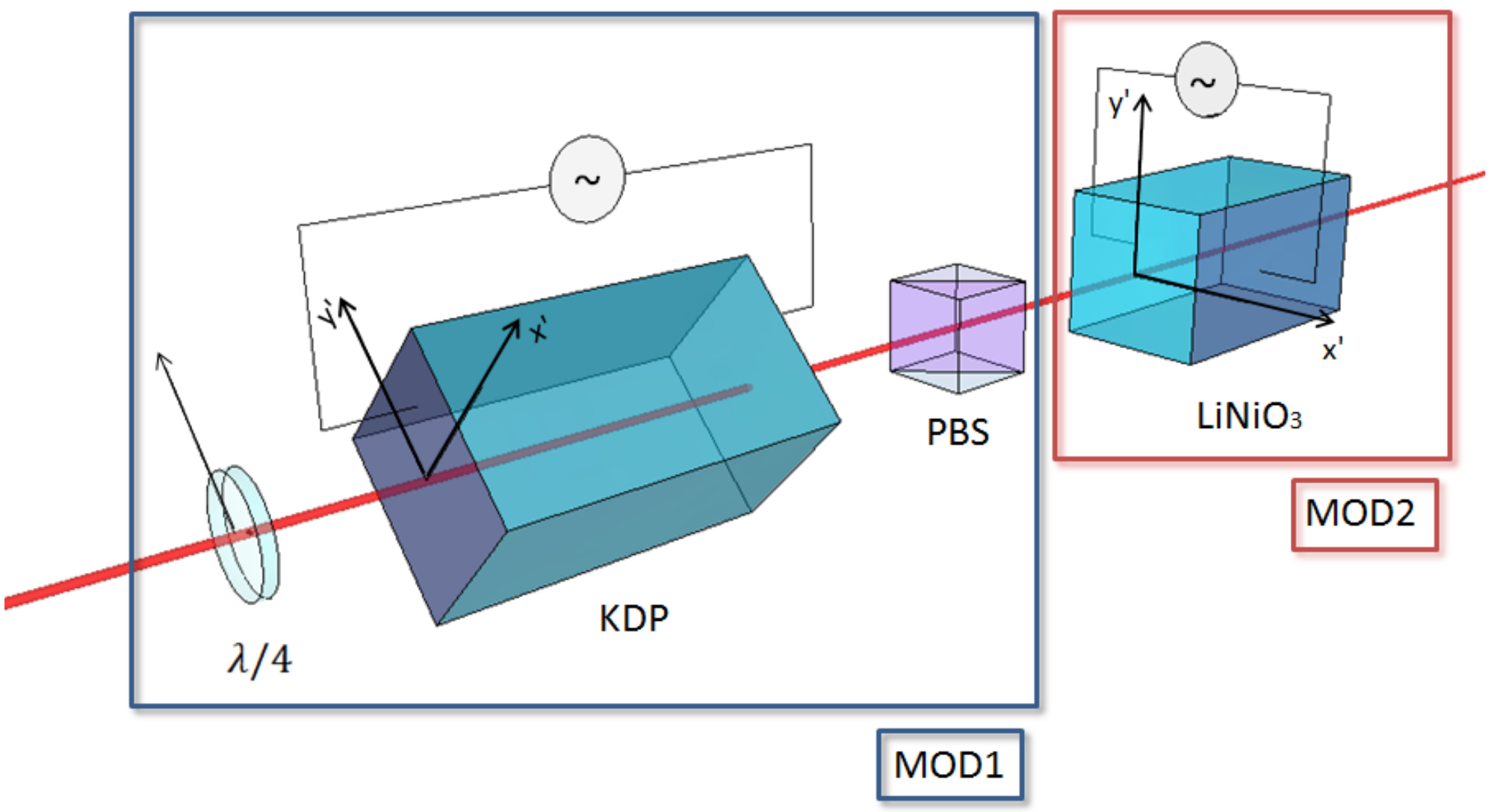}}
\caption{(Color online) (a) Schematic diagram of the experimental setup 
to generate squeezed thermal states. 
See text for details. (b) Enlarged picture of the optical systems
MOD1 and MOD2, used to generate the OPO input signals.  The optical
field is prepared with circular polarization by setting the fast axis of
$\lambda$/4 at an angle of $45^\circ$ with respect to the incident
\textit{p}-polarization, and then is passed through a KDP crystal whose
axis are oriented at $45^\circ$. The PBS select only the horizontal
component of output beam which is sent in a LiNiO$_3$ crystal whose
extraordinary axis is horizontal. } \label{f:schema}
\end{figure}
In this section we deal with the generation and the characterization 
of squeezed thermal states (STS) of a single-mode radiation field,
i.e. states of the form
\begin{equation}\label{STS}
\hat{\rho}= \hat{S}(r)\hat{\nu} (n_{\text{th}})\hat{S}^\dag(r)\,,
\end{equation}
where $\hat{S}(r)=\exp \big \{ \frac12 r \big 
[(\hat{a}^{\dag})^2 - \hat{a}^2 \big ] \big \}$ is the squeezing
operator, with $r\in\mathbb{R}$, $\hat{\nu} (n_{\text{th}})= 
n_{\text{th}}^{\hat{a}^\dag \hat{a}}/(1+n_{\text{th}})^{\hat{a}^\dag
\hat{a}+1}$ is a thermal state with $n_{\text{th}}$ average
number of photons and $[\hat{a},\hat{a}^\dag]=1$, $\hat{a}$ 
and $\hat{a}^\dag$ being field operators. Upon defining the 
quadrature operators
\begin{equation}\label{Quadrature}
\hat{x}_{\theta} \equiv \hat{a}\, {\rm e}^{- i \theta } 
+ \hat{a}^\dag \, {\rm e}^{ i \theta }\,,
\end{equation}
with $\theta \in [0,\pi]$, the STS are fully 
characterized by their first and second moments
\begin{subequations}\begin{align}
\label{Quad_ave}\langle \hat{x}_\theta \rangle &= 0 \qquad \forall \theta\\
\label{Quad_var} \langle \Delta \hat{x}_\theta^2 \rangle &= (1 + 2
n_{\text{th}})({\rm e}^{2r}\cos^2\theta + {\rm e}^{-2r}\sin^2\theta)\,,
\end{align}
\end{subequations}
where $\langle\cdots\rangle\equiv \hbox{Tr}[\hat\rho\,\cdots]$.
In terms of the canonical operators $\hat{x}\equiv\hat{x}_0$ and
$\hat{p}\equiv\hat{x}_{\pi/2}$, the covariance matrix (CM) of 
a STS reads
\begin{equation}\label{CM}
\sigma=\begin{pmatrix}
\langle \Delta \hat{x}^2 \rangle & 0 \\ 0 &  \langle \Delta \hat{p}^2 \rangle
\end{pmatrix}=\begin{pmatrix} s/\mu & 0\\ 
0 & 1/\mu s
\end{pmatrix},
\end{equation}
where $\mu=
{\rm Tr}[\hat{\rho}^2]= (2n_{\text{th}}+1)^{-1}$ is the 
purity of the state $\hat{\rho}$ and 
$s\equiv {\rm e}^{2r}$ is the squeezing factor. 
A STS is nonclassical, i.e. it corresponds to a singular Glauber 
P-function, whenever the conditions $s<\mu$ or $s > \mu^{-1} $ 
are satisfied.  The total energy of a STS is given by
\begin{equation}\label{Ntot}
N_{\text{tot}}=\langle \hat{a}^\dag \hat{a}\rangle =n_{\text{th}}+n_{\text{s}}+2 n_{\text{th}} n_{\text{s}}\quad ,
\end{equation}
where $n_s=\sinh^2 r$ is the number of squeezing photons 
and $n_{\text{th}}$ is the thermal contribution to energy.
\par
According to Eq.~(\ref{Ntot}), it is possible to find a suitable
parametrization of the single-mode STS CM (\ref{CM}) in terms of the
different energy contributions
\begin{subequations}\label{FitVariance}\begin{align}
\langle \Delta \hat{x}^2 \rangle&= \left(1 + 2 \frac{N_{\text{tot}} - n_{\rm s}}{2 n_{\rm s} + 1 }\right) (1 + 2 n_s - 2 \sqrt{n_{\rm s} + n_{\rm s}^2})\\
\langle \Delta \hat{p}^2 \rangle&= \left(1 + 2 \frac{N_{\text{tot}} - n_{\rm s}}{2 n_{\rm s} + 1 }\right)\frac{1}{(1 + 2 n_{\rm s} - 2 \sqrt{n_{\rm s} + n_{\rm s}^2})},
\end{align}\end{subequations}
from which the linear behavior of the variances as a 
function of the total energy $N_{\text{tot}}$ is apparent.
\par
The fidelity between two STS is given by
\cite{MarianFidelity}
\begin{equation}\label{FidGaussian}
F(\sigma_1, \sigma_2) =\frac{1}{\sqrt{\Delta + \delta} - \sqrt{\delta}},
\end{equation}
where $\Delta = \frac{1}{4} \det[\sigma_1 + \sigma_2 ]$ and $\delta=
\frac{1}{4} \prod_{i=1,2} (\det \sigma_i - 1)$. 
\subsection{Experimental setup}
\begin{table*}[t!]
\caption{\label{t:StateExp} Characterization, via homodyne tomography,
of the $m=14$ experimental STS in terms of the position and momentum
variances, total energy, squeezing factor and purity. The STS display
squeezing in position and anti-squeezing in momentum coordinates
($r<0$) .} \vspace{2mm}
\begin{ruledtabular}
\begin{tabular}{c | c  c  c  c  c} 
state \#& $ \langle \Delta \hat{x}^2 \rangle$ &  $\langle \Delta \hat{p}^2 \rangle$ & $\langle  \hat{a}^\dag \hat{a} \rangle$ & $s_{\text{exp}}$ & $\mu_{\text{exp}}$ \\[1mm]
\hline 
1 & $0.48\pm 0.03$ & $3.15\pm 0.09$ & $0.41\pm 0.02$ & $0.39\pm0.01$ & $0.81\pm0.03$   \\[.8mm]
2 & $0.67\pm 0.04$ & $3.33\pm 0.09$ & $0.50\pm 0.02$ & $0.45\pm0.01$  & $ 0.67\pm0.02 $  \\[.8mm]
3 & $0.62\pm 0.04$ & $3.77\pm 0.11$ & $0.60\pm 0.02$ & $0.40\pm0.02$  & $0.66\pm0.02$  \\[.8mm]
4 & $0.69\pm 0.05$ & $3.94\pm 0.11$ & $0.66\pm 0.02$ & $ 0.41\pm 0.02$ & $ 0.61\pm0.02 $  \\[.8mm]
5 & $0.70\pm 0.05$ & $4.51\pm 0.12$ & $0.80\pm 0.03$ &$ 0.39\pm 0.02$ & $ 0.56\pm0.02 $   \\[.8mm]
6 & $0.77\pm 0.05$ & $4.54\pm 0.13$ & $0.83\pm 0.03$ & $ 0.41\pm 0.02 $ & $ 0.54\pm0.02 $   \\[.8mm]
7 & $0.77\pm 0.05$ & $4.60\pm 0.13$ & $0.84\pm 0.03$ &  $ 0.41\pm 0.02 $ & $ 0.53\pm0.02 $   \\[.8mm]
8 & $0.93\pm 0.06$ & $5.00\pm 0.14$ & $0.98\pm 0.03$ &  $ 0.43\pm 0.02 $ & $ 0.46\pm0.02 $    \\[.8mm]
9 & $0.95\pm 0.06$ & $5.36\pm 0.15$ & $1.08\pm 0.03$ &  $ 0.42\pm 0.01 $ & $ 0.44\pm0.02 $   \\[.8mm]
10 & $0.93\pm 0.07$ & $5.56\pm 0.15$ & $1.12\pm 0.03$ &  $0.41\pm 0.02$ & $ 0.44\pm0.02 $  \\[.8mm]
11 & $1.00\pm 0.07$ & $5.80\pm 0.17$ & $1.20\pm 0.03$ & $ 0.42\pm 0.02$ & $ 0.42\pm0.02 $  \\[.8mm]
12 & $1.13\pm 0.07$ & $5.87\pm 0.16$ & $1.25\pm 0.03$ & $ 0.44\pm 0.02 $ & $ 0.39\pm0.01 $    \\[.8mm]
13 & $1.11\pm 0.08$ & $6.33\pm 0.18$ & $1.36\pm 0.04$ &  $ 0.42\pm 0.02 $ & $ 0.38\pm0.01 $    \\[.8mm]
14 & $1.30\pm 0.08$ & $6.16\pm 0.18$ & $1.36\pm 0.04$ &  $ 0.46\pm 0.02 $ & $ 0.35\pm0.01 $  \\[.8mm]
\end{tabular}
\end{ruledtabular}
\end{table*}
In order to generate STS we employ the experimental setup
schematically depicted in Fig.~\ref{f:schema} (a).  It consists of
three stages: Laser, signal generator (SG) and homodyne detector (HD).
Our source is a home-made internally frequency doubled Nd:YAG laser. It
is based on a 4 mirrors ring cavity and the active medium is a
cylindrical Nd:YAG crystal (diameter 2 mm and length 60 mm) radially
pumped by three array of water-cooled laser diodes @ 808 nm. The crystal
for the frequency doubling is a periodically poled MgO:LiNbO$_3$ (PPLN)
of 10 mm thermally stabilized ($\sim$70$^\circ$C). Inside the cavity is
placed a light diode that consist of a half-wave-plate (HWP), a Faraday
rotator (15$^\circ$) and a Brewster plate (BP) in order to obtain the
single mode operation.  
\par
The laser output $@$ 532~nm is used as the pump for a optical parametric
oscillator (OPO) while the output at 1064~nm is split into two beams by
using a polarizing beam splitter (PBS): one is used as the local
oscillator (LO) for the homodyne detector and the other as the input for
the OPO. The OPO cavity is linear with a free spectral range (FSR) of
3300~MHz, the output mirror has a reflectivity of 92\% and the rear
mirror 99\%.  A phase modulator (PM) generate a signal at frequency of
110~MHz (HF) used as active stabilization of the OPO cavity via 
the Pound-Drever-Hole (PDH) 
technique \cite{PDH,SBrec}:  the reflected beam from cavity is detected (D)
and used to generate the error signal of PDH apparatus. This signal
error drives a piezo connected to the rear mirror of the OPO cavity to
actively control its length.
\par  
The homodyne detector (HD) consists of a 50:50 beam splitter,  two low
noise detector and a differential amplifier based on a LMH6624
operational amplifier. The visibility of the interferometer is about
98\%.  To remove the low frequency signal we use an high-pass filter
@~500~kHz and then the signal is sent to the demodulation apparatus. The
information about the signal, that is at frequency $\Omega$ about 3MHz,
is retrieved by using an electronic apparatus which consists of a phase
shifter a mixer and a low pass filter @~300~kHz. The LO phase is spanned
between 0 and $2\pi$ thanks to a piezo-mounted mirror linearly driven by
a ramp generator (RG).  
\par
Our goal is to study a single-mode squeezed thermal state and therefore
we have to generate a thermal seed to be injected into the OPO.
The density matrix of thermal state in the Glauber
representation reads as follows 
\begin{equation}\label{State_exp}
\hat{\nu}_{\text{exp}}(n_{\rm th})=\int_0^\infty d\lvert\alpha\lvert 
\frac{2|\alpha|}{n_{\rm th}}e^{-\frac{\lvert\alpha\lvert^2}{n_{\rm th}}}
\int_0^{2\pi}\frac{d\phi}{2\pi}\lvert\lvert\alpha\lvert e^{i\phi}
\rangle  \langle\lvert\alpha\lvert e^{i\phi}\lvert\,,
\end{equation}
i.e., it can be viewed as a mixture of coherent states with phase $\phi$
uniformly distributed over the range $0$ to $2\pi$, and a given amplitude
$\lvert\alpha\lvert$ distribution. Therefore, we have to generate a
rapid sequence of coherent states with $\lvert\alpha\lvert$ and $\phi$
randomly selected from these distributions. 
\par
Our strategy is to exploit the combined effect of the two optical
systems (MOD1 and MOD2 in Fig.~\ref{f:schema} (a) described in
Ref.~\cite{bachor}  and sketched in more detail in Fig.~\ref{f:schema}
(b).  MOD1 generate a coherent state with phase $0$, while MOD2 generate
a coherent state with phase $\frac{\pi}{2}$. By matching these coherent
states with properly chosen amplitudes, it is possible to generate an
arbitrary coherent state. In order to control this process via pc, the
MOD1 and MOD2 are driven by two identical electronic circuits which
consist of a phase shifter and a mixer. The pc processes the
$\lvert\alpha\lvert$ and $\phi$ values of the coherent state which we
want to generate, and convert them into the voltage signals which are sent
to the mixer together with the sinusoidally varying signals at frequency
$\Omega$ in order to obtain the right modulation signals on MOD1 and
MOD2.  
\par
Finally, in order to obtain the desired thermal state, the pc generates
random $\lvert\alpha\lvert$ and $\phi$ values according to their
specified distributions (see Eq. (\ref{State_exp})) and converts them in
two simultaneous trains of voltage values which are sent to the crystals
in a time window of 70 ms with a repetition rate of 100 kHz. Generation
and acquisition operations are synchronized in the same time window at the same sampling rate.
Therefore we collect 7000 homodyne data points $\{(\theta_k,x_k)\}$, 
LO phase and quadrature value, respectively. 
The sampling is triggered by a signal generated by RG to ensure the
synchronization between the acquisition process and the scanning of LO
with $\theta_k \in [0,2\pi]$.
\par
Notice that seeding the OPO is a crucial step to observe the
quantum-to-classical transition with STS. As a matter of 
fact, without seeding the OPO, output signal is a squeezed 
vacuum state, which is then degraded to a STS with a nonzero 
thermal component by propagation in a lossy channel. However, 
STS obtained in this way are always non-classical
for any value of the loss and the squeezing 
parameters \cite{rossi:04,fop:05,oli:rev}. 
\subsection{Homodyne Tomography}
\begin{figure}
\includegraphics[width=0.9\columnwidth]{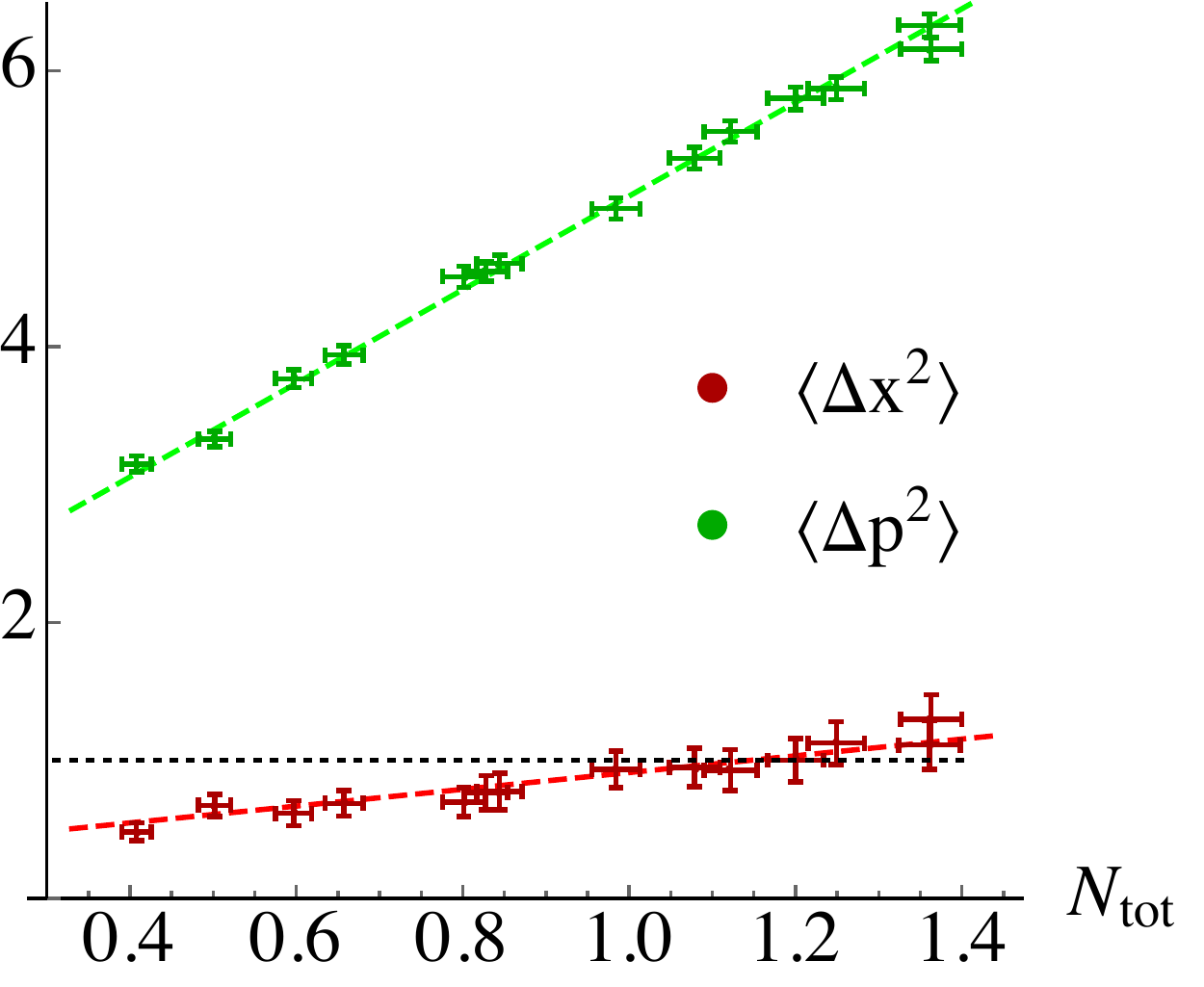}
\caption{(Color online) Tomographic reconstruction of the variances of
the squeezed quadrature $\hat{x}$ (red dots) and of the anti-squeezed
quadrature $\hat{p}$ (green dots) as a function of the total energy
$N_{\text{tot}}$, for $m=14$ experimental STS. Dashed lines represent
linear fits of the experimental data (see Eqs. (\ref{FitVariance})),
from which we obtain the number of squeezed photons $n_{\rm s} \simeq 0.2$. The
black dotted horizontal line is the shot-noise level at $ \langle \Delta
\hat{x}^2 \rangle= \langle \Delta \hat{p}^2 \rangle=1$.}
\label{f:ThSqVariance}
\end{figure}
We perform state reconstruction of single-mode CV systems 
by quantum homodyne tomography, i.e. by collecting homodyne data 
at different LO phases and applying the pattern functions 
method \cite{revt1}. This technique allows one to obtain the 
expectation value of any observable $\hat{O}$ on a given state $\hat{\rho}$ 
starting from a set of homodyne data  $\{(\theta_k , x_k)\}$,
$x_k$ being the $k$-th outcome from the 
measurement of the quadrature (\ref{Quadrature}) at phase $\theta_k$, 
with $k=1,\ldots,M$. Upon exploiting the Glauber representation
of operators in polar coordinates, the average value of a generic
observable $\hat{O}$ may be rewritten as 
\begin{equation}\label{O_ave}
\ave{\hat{O}}= \int_0^\pi \frac{d\theta}{\pi} 
\int_{-\infty}^{+\infty} dx\, p(x, \theta) 
\mathcal{R}[\hat{O}] (x, \theta),
\end{equation} 
where $p(x, \theta)= \bra{x_\theta} \hat{\rho} \ket{x_\theta}$ is the 
distribution of quadrature outcomes, with 
$\{\ket{x_\theta}\}$ the set of eigenvectors of $\hat{x}_\theta$, and 
$\mathcal{R}[\hat{O}] (x, \theta )= \int_{-\infty}^{+\infty} dy |y| 
\Tr[\hat{O}e^{i y (\hat{x}_\theta- x)}]$ is the estimator of 
the operator ensemble average $\ave{\hat{O}}$. For large samples 
$M\gg 1$, the integral (\ref{O_ave}) can be recast in the 
discrete form
\begin{equation}\label{O_ave_discrete}
\ave{\hat{O}}\simeq \frac{1}{M}\sum_{k=1}^M  
\mathcal{R}[\hat{O}] (x_k, \theta_k)\,.
\end{equation}
The uncertainty of the estimated value $\ave{\hat{O}}$ 
is ruled by the central limit theorem and scales as $\sqrt{M}$, namely
\begin{equation}\label{Precision_O}
\delta \langle \hat{O}
\rangle=\frac{1}{\sqrt{M}}\sqrt{\sum_{k=1}^M\frac{\big[\mathcal{R}[\hat{O}]
(x_k, \theta_k)\big]^2 - \langle \hat{O} \rangle^2}{M}}.  \end{equation}
In order to properly characterize a single-mode prepared in a 
Gaussian STS, we need to estimate the first two moments of 
the quadrature operator $\hat{x}_\phi$ and reconstruct the 
first-moment vector and the CM, as well as the total energy 
$\hat{a}^\dag \hat{a}$ of the state. We thus need 
the following estimators \cite{revt1}:
\begin{subequations}\begin{align}
\mathcal{R}[\hat{x}_\phi] (x, \theta) &= 2 x \cos(\theta-\phi)\\
\mathcal{R}[\hat{x}_\phi^2] (x, \theta) &= (x^2-1)\Big\{1+2\cos[2(\theta-\phi)]\Big \}+1\\
\mathcal{R}[\hat{a}^\dag \hat{a}] (x, \theta) &=\frac12 \left (x^2-1\right ).
\end{align}\end{subequations}
In this way it is possible to compute the average value $\langle \hat{O}
\rangle$ and the fluctuations $\langle \Delta\hat{O}^2 \rangle\equiv \langle
\hat{O}^2 \rangle-\langle \hat{O} \rangle^2$ for the observables of
interest, toghether with the corresponding uncertainties (\ref{Precision_O}).
\par
We collect $M=7000$ homodyne data $\{(x_k, \theta_k)\}$ for each state
and address the quantum-to-classical transition by generating $m=14$
STS with increasing thermal component, as
the squeezing is fixed by the geometry of the experimental setup. For
all the generated states, we tested the compatibility with the typical
form of the STS, i.~e. null first-moment vector (\ref{Quad_ave}) and
diagonal CM (\ref{CM}). We characterized these states (see Table
\ref{t:StateExp}) in terms of the position $\langle \Delta \hat{x}^2
\rangle$ and momentum $ \langle \Delta \hat{p}^2 \rangle$ variances, the
total energy $N_{\text{tot}}\equiv\langle \hat{a}^\dag \hat{a} \rangle$,
together with the squeezing factor $s_{\text{exp}}$ and the purity
$\mu_{\text{exp}}$. As already mentioned at the beginning of this
section, the shot-noise threshold is set at $ \langle \Delta \hat{x}^2
\rangle= \langle \Delta \hat{p}^2 \rangle=1$, under which the state of
the detected single-mode radiation displays genuine quantum squeezing.
The generated STS display squeezing in position quadrature and
anti-squeezing in momentum quadrature (i.e. we have real and negative
squezing parameter $r<0$). In
Fig.~\ref{f:ThSqVariance} we show the position and momentum variances as
a function of the total energy for the $m=14$ experimentally generated
STS.  A linear fitting, following Eq. (\ref{FitVariance}),
provides the value of the number of squeezed photons $n_{\rm s}\simeq 0.2$,
which corresponds to $\sim 3.7~ \text{dB}$ of squeezing. 
Figure~\ref{f:ThSqVariance} makes apparent
the  capability of the experimental setup to generate
STS-on-demand by seeding the OPO with a controlled number
of thermal photons and in turn, to monitor the quantum-to-classical 
transition of a single-mode Gaussian state of light.
\subsection{Fidelity}
\label{s:CVfidelity}
\begin{figure}[h!]
\includegraphics[width=0.99\columnwidth]{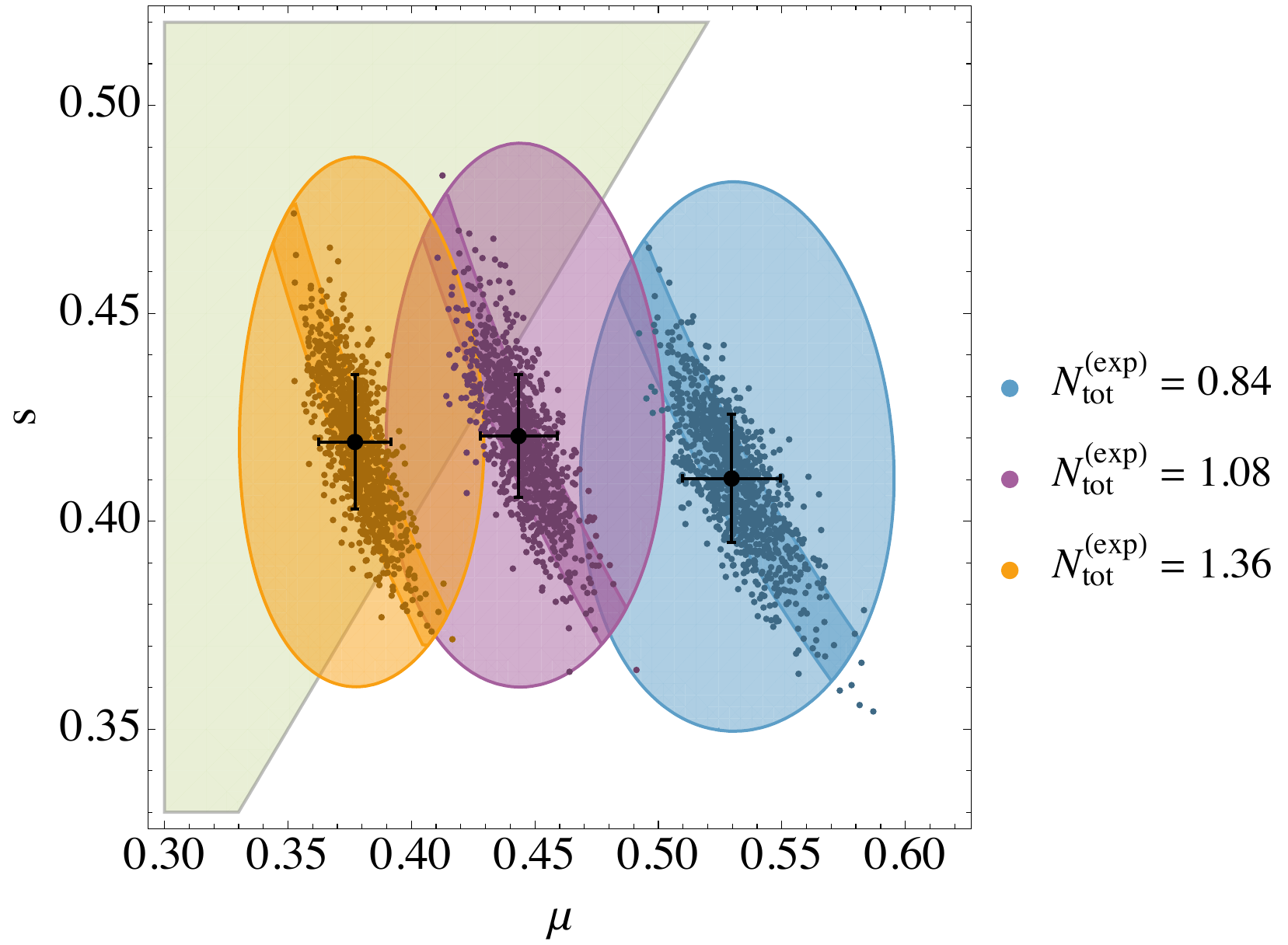}
\caption{(Color Online) Statistical distribution of reconstructed
STS in the squeezing-purity plane-$\{s,\mu\}$. Data come from 
$N_{\text{MC}}$ Monte Carlo resampled set of data for STS (see 
text and Appendix \ref{s:MC}). From left to right, we have distributions 
for  three experimental STS (state number $7, 9$ and $13$ of 
Table~\ref{t:StateExp}), shown as black
points with the corresponding bars of precision. The triangular-like
region $s>\mu$ contains states with a classical analogue.
The whole set of reconstructed states are contained in the ovoidal 
regions, i.e. have fidelity $F>0.995$ to the corresponding target 
state. The stripe-like regions are obtained adding a constraint to
the total energy, i.e. 
$N_{\text{tot}}^{(\text{exp})}-\delta N_{\text{tot}}^{(\text{exp})} <
\langle \hat{a}^\dag\hat{a} \rangle <
N_{\text{tot}}^{(\text{exp})}+\delta N_{\text{tot}}^{(\text{exp})}$.}
\label{f:ThsqBorder}
\end{figure}

\begin{figure}[h!]
\includegraphics[width=0.99\columnwidth]{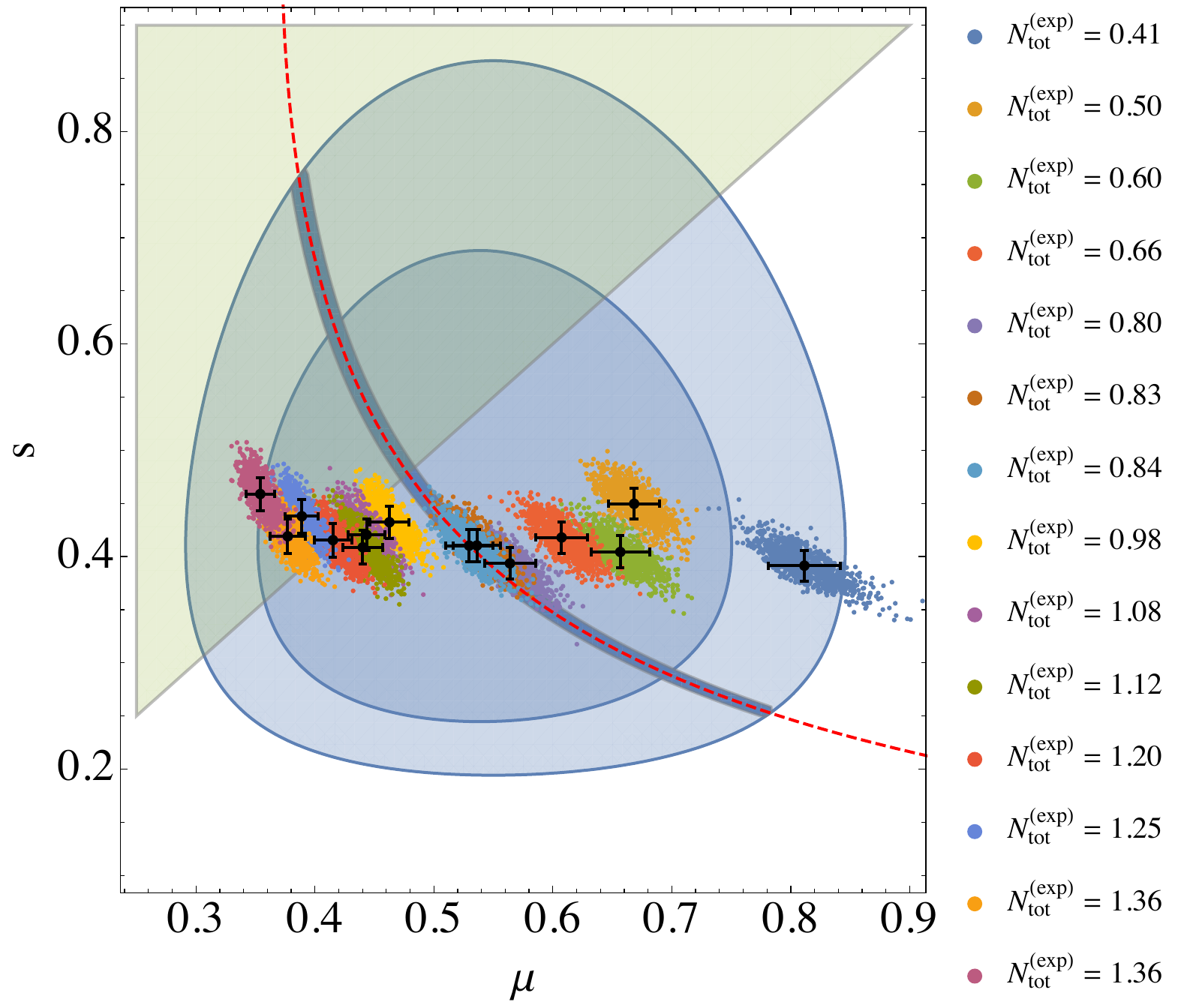}
\caption{(Color online) Statistical distribution of reconstructed
STS in the squeezing-purity plane-$\{s,\mu\}$ for all the experimental
target states in Table \ref{t:StateExp} (black points). The two balloons
include states  having fidelity to a nonclassical target STS
(with $s=0.41$ and $\mu=0.53$) larger than $F>0.90$ (outer balloon) or
$F>0.95$ (inner balloon) respectively, i.e. values commonly recognized 
as regions of {\em high fidelity}.  The size of the compatibility 
regions may be reduced by adding energy constraints (as discussed 
in Fig.~\ref{f:ThsqBorder}). A significant amount of states may still 
display opposite classicality properties compared to the target.} 
\label{f:ThSq09}
\end{figure}

In order to perform the uncertainties budget, to discuss the 
statistical distribution of relevant quantities, and to assess 
the statistical significance of fidelity, we generate 
$N_{\text{MC}}=10^3$ Monte Carlo replica data samples 
(see Appendix \ref{s:MC}), for each experimental state. 
Resampled (raw) homodyne data are drawn from Gaussian distributions
using the experimental values of Table~\ref{t:StateExp} to build
the average values (\ref{Quad_ave}) and the variances 
(\ref{Quad_var}) of the distributions.
For all the $m=14$ STS we apply homodyne tomography and analyze the 
distribution of the reconstructed states in the neighbouring of the 
experimental target state. 
Results are shown in Fig.~\ref{f:ThsqBorder} and Fig.~\ref{f:ThSq09}. 
using the squeezing-purity plane-$\{s,\mu\}$ representation.
Figure~\ref{f:ThsqBorder} focuses on three specific states (number 7, 9
and 13 of Table \ref{t:StateExp}) which are closer to the
quantum-classical boundary. Target states correspond to black points
whereas the ovoidal regions denotes the set of states having fidelity
larger than $F>0.995$ to the target. The darker, stripe-like, 
regions within each balloon correspond to states satisfying
the additional constraint of having fluctuations of 
the total energy (\ref{Ntot}) at most within one standard 
deviation.
\par
As it is apparent from the plot, the distribution of STS 
is concentrated within those stripes. On the other hand, 
despite the distributions are very sharp in terms
of fidelity to their targets
(remind that the balloons contains states with fidelity
larger than $F\geq 0.995$ to the target), their physical 
properties may be very different. This fact is clearly 
illustrated looking at nonclassicality, i.e. checking 
whether the Glauber $P$-function of the state is regular 
(this happens if $s>\mu$, corresponding to a triangular 
region in Fig. \ref{f:ThsqBorder}) or singular: states 
with very high fidelity to a classical or
a nonclassical target may not share this property with the 
target itself.
\par
This effect may be not particularly surprising for target states at border 
of the classicality region, even for high values of fidelity. On the 
other hand, the point becomes far more relevant if values of
fidelity commonly used in experiments are considered. In
Fig.~\ref{f:ThSq09} we show the balloons of states having
fidelity $F\geq0.9$ or $F\geq0.95$ to a nonclassical target
STS. As it is apparent from the plot, {\em all} the generated
STS are contained in the ballons, irrespectively of their nonclassicality.
The compatibility region may be considerably reduced in size by adding 
an energy constraint but, nonetheless, a large number of states may 
still fall in the region of classicality. 
\par
Overall, we conclude that fidelity is not a significant figure of
merit to assess nonclassicality of STS and should not be employed
to benchmark a generation scheme or certify quantum resources 
for a given protocol.
\section{Two-qubit systems}
\label{s:Werner}
In this Section we deal with discrete two-qubit systems.
In particular we focus on two-photon polarization states 
$\ket{HH}$, $\ket{HV}$, $\ket{VH}$ and $\ket{VV}$, and address
the reconstruction of statistical mixtures belonging to the class 
of Werner states: \begin{equation}\label{Werner}
\hat{\rho}^{(w)}=p\,|\Psi^-\rangle\langle \Psi^-|
+\frac{1-p}{4} \hat{\mathbb{I}}_4,
\end{equation}
where $\hat{\mathbb{I}}_4$ is the identity operator 
in the 4-dimensional Hilbert space of two qubits 
and $\ket{\Psi^-}$ is one of the maximally entangled Bell states
\begin{equation}\begin{split}
\ket{\Phi^\pm}=\frac{\ket{HH} \pm\ket{VV}}{\sqrt{2}} \quad\mbox{and}\quad
\ket{\Psi^\pm}=\frac{\ket{HV} \pm\ket{VH}}{\sqrt{2}}.
\end{split}\end{equation}
The parameter $-1/3\leq p\leq 1$ tunes the mixture (\ref{Werner}) from
the maximally mixed state $\hat{\mathbb{I}}_4/4$ for $p=0$ to the
maximally entangled Bell state $\ket{\Psi^-}$ for $p=1$. In between, the
quantum-to-classical transition is located at $p =1/3$, with 
entangled states satisfying $p>1/3$ and separable ones 
$p<1/3$. 
\par
\begin{figure}[h!]
\includegraphics[width=0.95\columnwidth]{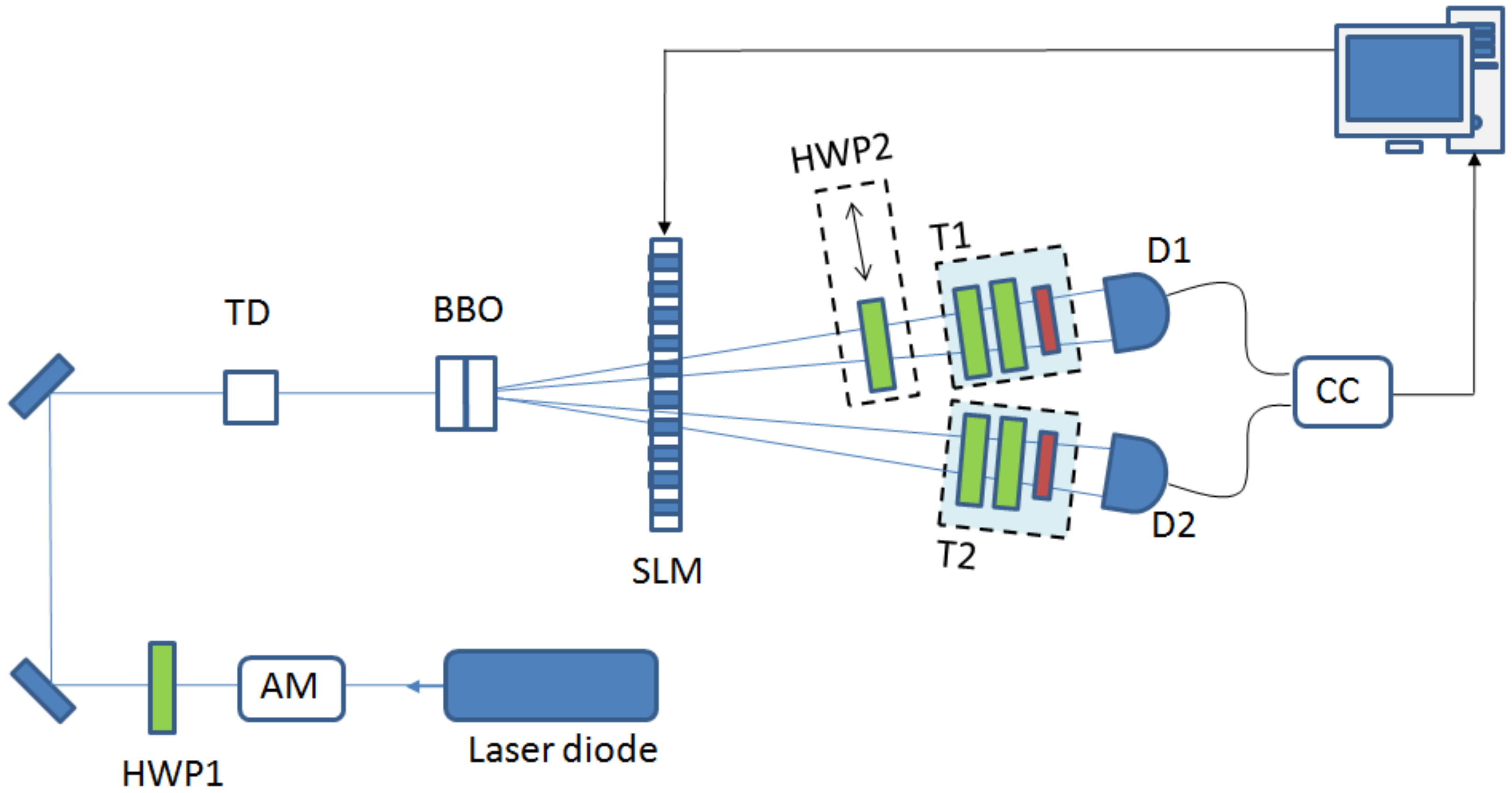}
\caption{(Color online) Schematic diagram of experimental setup.
A linearly polarized cw laser diode at 405 nm pumps a couple of BBO
crystals cut for type-I downconversion. The horizontal and vertical
amplitudes of the photon pairs are balanced by a half-wave plate set
along the pump path (HWP1), whereas an additional BBO crystal (TD) is
placed on the pump path to compensate the temporal delay. The amplitude
modulator (AM) consists of a half-wave-plate and
polarizer-beam-splitter. Signal and idler beams travel through the SLM,
which provides purification of the generated states. A half-wave plate
(HWP2) is inserted on signal path in order to generate the state
$\hat{\rho}_\lambda$ (see the text), whereas a quarter-wave plate, a
half-wave plate, and a polarizer (sectors T1 and T2)  are used for the
tomographic reconstruction. Finally the beams are detected by detectors
D1 and D2 ad sent to single-photon counting modules (CC)}
\label{f:discreto}
\end{figure}
\subsection{Experimental generation of Werner states}
A schematic diagram of experimental setup is sketched in
Fig.~\ref{f:discreto}. Photon pairs are generated by
type-I downconversion from a couple of beta barium borate (BBO)
crystals, in a non-collinear configuration, pumped with a linearly
polarized cw 405 nm laser diode, whose effective power on the generating
crystals is about 10 mW. The experimental apparatus has been already
described in detail in Refs.~\cite{generazione,generazione2}. Here a
half-wave plate (HWP2 in Fig.~\ref{f:discreto}) has been 
inserted in front of detector D1 to perform 
$\ket{\Phi^-} \rightarrow \ket{\Psi^-} $
transformation. A programmable one-dimensional spatial light modulator
(SLM) is placed on the path of signal and idler in order to control the
visibility of the generated states. The SLM provides the setup with
great flexibility, allowing the experimenter to choose and set the
visibility of generated states \cite{SLM,SLM2}. Eventually, photons are
focused in two multimode fibers and sent to single-photon counting
modules (CC).  
\par 
Our experimental apparatus allows us to mix two
types of Bell states at a time, either $\ket{\Psi^\pm}$ or
$\ket{\Phi^\pm}$. In order to obtain a Werner state (\ref{Werner}) we
generate the polarization entangled states $\hat{\rho}_\lambda=
\lambda\ketbra{\Psi^-}{\Psi^-}+(1-\lambda)\ketbra{\Psi^+}{\Psi^+}$ and
the mixed state $\hat{\rho}_{\rm mix}=\left(
\ketbra{\Phi^+}{\Phi^+}+\ketbra{\Phi^-}{\Phi^-} \right)/2 $
\cite{generazione,visibilita}. Werner states may be obtained 
by suitably mixing these two states
$\hat{\rho}^{(w)}=f_1\hat{\rho}_\lambda+f_2\hat{\rho}_{\rm mix}$ with
proper probabilities, given by $f_1=\frac{1+p}{2}$,  $f_2=\frac{1-p}{2}$
and $\lambda=\frac{2p}{p+1}$. The mixed state $\hat{\rho}_{\rm mix}$ is
obtained using the same scheme of Fig.~\ref{f:discreto} upon removing the
HWP2 from the signal path and setting the SLM in order to get $\lambda\simeq
0$. The frequencies $f_1$ and $f_2$ are tuned by changing the power of the
pump beam with an amplitude modulator (AM). The full range of Werner
states may be explored. 
\par
 \begin{table*}[t!]
\caption{\label{t:Werner} Statistical analysis of the tomography of
$N_{\text{MC}}=10^3$ two-qubit states, having fidelities
$F(\overline{\hat{\rho}_k},\hat{\rho}_k^{(w)})$ with target Werner of
parameter $p_{k}^{(w)}$. The average values of the least eigenvalue
$e_m(\hat{\rho}^{(\tau)})$ and of quantum discord $D(\hat{\rho})$ are
reported for both the distributions of tomographic states and of the
approximated Werner states. } \vspace{2mm}
\begin{ruledtabular}
\begin{tabular}{c | c  c  c  c } 
 &  state 1 &  state 2 & state 3 & state 4 \\[1mm]
\hline 
$p_{k}^{(w)}$ & $0.32\pm0.04$ & $0.35\pm0.04$ & $0.28\pm0.04$ & $0.44\pm0.05$ \\[.8mm]
$F(\overline{\hat{\rho}_k},\hat{\rho}_k^{(w)})$ & $0.985^{+ 0.006}_{- 0.01}$ & $0.988^{+ 0.005}_{- 0.01}$ & $0.987^{+ 0.006}_{- 0.01}$ & $0.985^{+ 0.007}_{- 0.02}$  \\[.8mm]
$e_m(\overline{\hat{\rho}_k}^{(\tau)})$ & $0.01\pm0.03$ & $-0.01\pm0.03$ & $0.04\pm0.03$ & $-0.07\pm 0.03$  \\[.8mm]
$e_m([\hat{\rho}_k^{(w)}]^{(\tau)})$ & $0.01\pm0.03$ & $-0.01\pm0.03$ & $0.04\pm0.03$  & $-0.08\pm 0.04$  \\[.8mm]
$D(\overline{\hat{\rho}_k})$ & $0.08\pm0.02$ & $0.10\pm0.02$ & $0.06\pm0.02$ & $0.14 \pm 0.02$  \\[.8mm]
$D(\hat{\rho}_k^{(w)})$ & $0.11\pm0.03$ & $0.14\pm0.03$ & $0.06\pm0.02$ & $0.21 \pm 0.04$  \\[.8mm]
\end{tabular}
\end{ruledtabular}
\end{table*}
The tomographic reconstruction is performed by measuring 16 projective
and independent observables in the two-qubit Hilbert space, namely
$P_j=\ketbra{\psi_j}{\psi_j}$ (with $j= 1, \ldots,16$). Different settings of
the apparatus, obtained by combining a quarter-wave plate, a half-wave
plate and a polarizer (sectors T1 and T2 in Fig.~\ref{f:discreto}), are
employed \cite{kb00,jam01}. Each of the 16
measurements correspond to 30 acquisitions, in a time window of $1~s$,
of coincidence photon counts $\{n_j \}_{j=1}^{16}$,  i.e. the outcomes
of the projectors $P_j$. 
\subsection{Tomography with MLE}
The density matrix of the two-qubit states generated in 
the experiment has been reconstructed using maximum-likelihood 
(MLE) tomographic protocol
\cite{kb00,jam01}. This scheme adopts a suitable
parametrization of the density matrix, namely $\hat{\rho}(\mathbb{T})=
T^{\dag}T / \Tr[T^{\dag}T]$, where $T$ is a complex lower triangular
matrix and $\mathbb{T}=\{t_j\}_{j=1}^{16}$ is the set of 16 parameters
characterizing the density matrix. In this way it is ensured that
$\hat{\rho}$ is positive and Hermitean (Choleski decomposition). The MLE
protocol allows to recover the set $\mathbb{T}$ by means of a
constrained optimization procedure with Lagrange multipliers, which
accounts for the normalization condition $\Tr[\hat{\rho}]=1$, involving
the set of data coming from the 16 experimental measurements. In
particular, the logarithmic likelihood functional to be minimized
reads
\begin{equation}
\mathcal{L}(\mathbb{T})= \sum _{j=1}^{16} \frac{ [\,\mathcal{N}\bra
{\psi_{j}} \hat{\rho}(\mathbb{T}) \ket {\psi_{j}} - n_j \,]^2}{2\,
\mathcal{N} \bra {\psi_{j}} \hat{\rho}(\mathbb{T}) \ket {\psi_{j}} },
\end{equation}
where $\mathcal{N} =  \sum _{j=1}^{4} n_j $ is a 
constant proportional to the total number of 
acquisitions. 
\par
We experimentally generated $N_{\text{exp}}=4$ two-qubit states 
not too far from the border between separable and entangled 
states (see Table~\ref{t:Werner}). MLE quantum tomography 
shows that the reconstructed density matrices do not 
display the typical X-shape of an ideal Werner
state (\ref{Werner}) with real-valued elements. A possible route to
extract the desired Werner state, is based on the maximization of the
fidelity between the experimental state and the generic Werner state
(\ref{Werner}). This procedure sounds reasonable and, in principle,
may allow one to assess the quantum resources contained in the generated
state, as well as to exploit them in order to
accomplish quantum tasks. On the other hand, we will show in the following 
that a fidelity-based inference is in general misleading and 
should be avoided in assessing the true quantum properties of the
experimentally generated state.
\subsection{Fidelity}
\label{s:fidelityqubit}
In order to perform statistical analysis of data and evaluate
uncertainties we have resampled 
photon counts data to obtain $N_{\text{MC}}=10^3$ repeated samples 
for each of the $N_{\text{exp}}=4$ experimental states 
(see Appendix \ref{s:MC} for details). 
The significance of fidelity may be assessed upon the 
comparison between two possible strategies to reconstruct 
quantum properties of the generated states and their 
distribution. In the first strategy we evaluate properties 
from the reconstructed states 
$\hat{\rho}_k^i$ ($k=1,\ldots,N_{\text{exp}}$ and
$i=1,\ldots,N_{\text{MC}}$) as obtained by the MLE tomography, 
whereas in the second one we look for the closest Werner 
state, in terms of fidelity, for each reconstructed state 
and analyze the properties of this class of Werner 
states $[\hat{\rho}_k^i]^{(w)}$. 
\par
The first method is based only 
in tomographic data and provides an average MLE two-qubit state
$\overline{\hat{\rho}_k}=\sum_{i=1}^{N_{\text{MC}}}\hat{\rho}
_k^i/N_{\text{MC}}$,
which optimize the likelihood of the experimental data. This 
average state may be then employed to infer a Werner target
state $\hat{\rho}^{(w)}_k$, of the form given in Eq. (\ref{Werner}), via
a maximization of the fidelity
$F(\overline{\hat{\rho}_k},\hat{\rho}_k^{(w)})$. Upon adopting the second
strategy, we obtain a distribution of approximated Werner states, with
an average state compatible, at least in principle, 
with the Werner target state of
the first strategy. The parameters $p_{k}^{(w)}$ characterizing the
Werner target states are reported in Table~\ref{t:Werner}.  
\par
In the following, we analyze how some properties of the quantum states
distribute around the target states in terms of fidelity, depending
on which of the two strategies has been adopted. In particular we
consider the amount of quantum correlations of two-qubit states,
as quantified by entanglement and quantum discord.  
\begin{figure}
\subfigure[]{\includegraphics[width=0.4\textwidth]{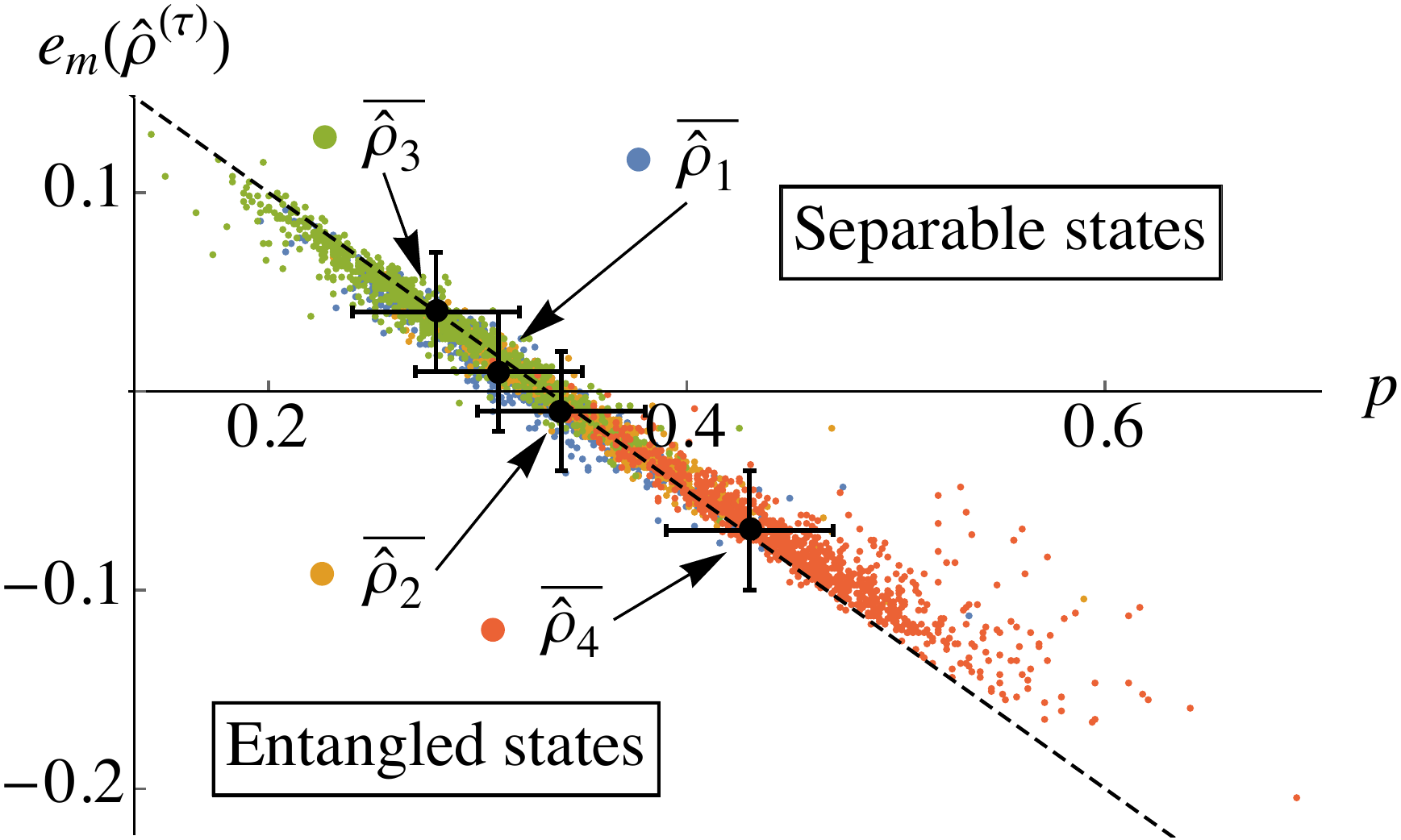}} 
\subfigure[]{\includegraphics[width=0.4\textwidth]{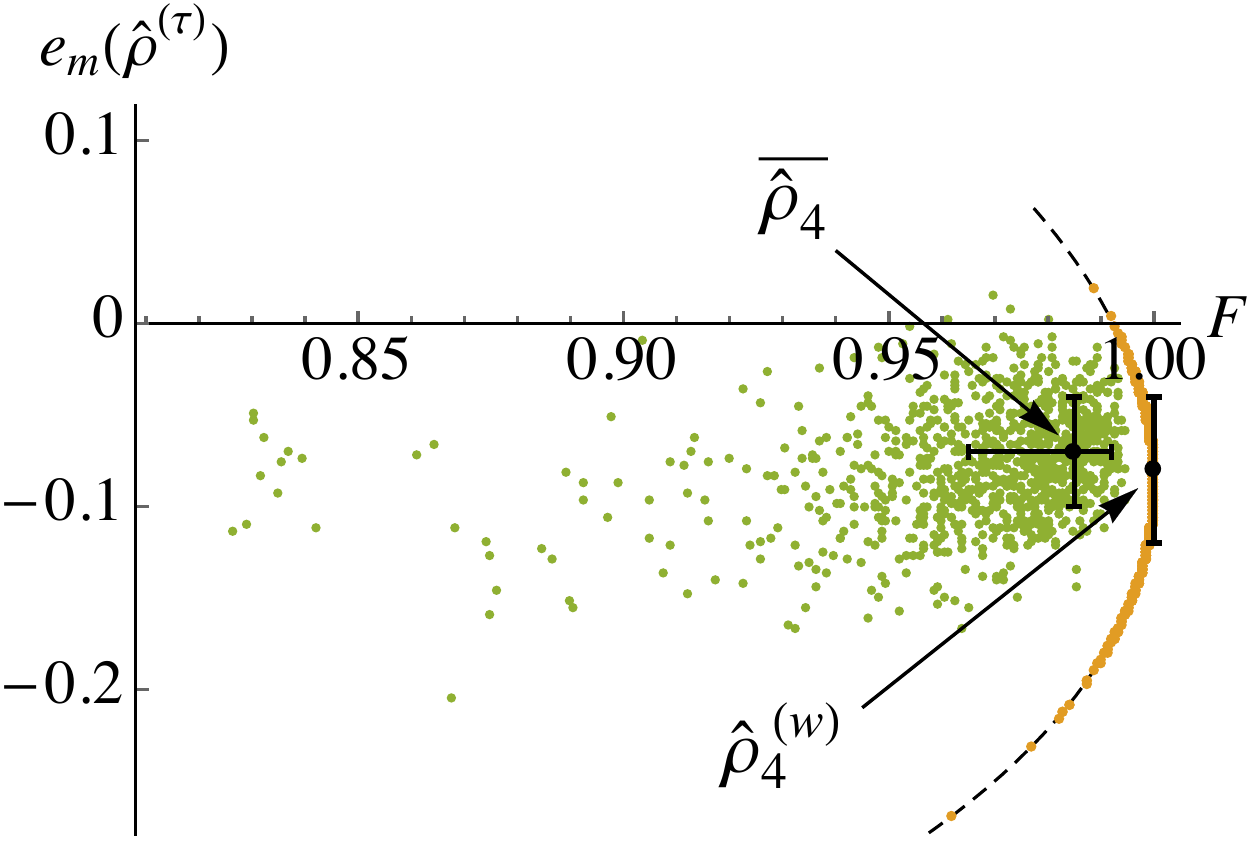}} 
\caption{(Color online) (a) Distribution of
$e_m([\hat{\rho}_k^i]^{(\tau)})$ for $10^3$ resampled states as a
function of the Werner parameter $p$. The $N_{\text{exp}}=4$ average
states $\overline{\hat{\rho}_k}$ are highlighted with black dots and
error bars, matching the theoretical curve
$e_m(\hat{\rho}^{(w)})=(1-3p)/4$ (dashed black line) relative to the
negative eigenvalue of the partially transposed ideal Werner state
(\ref{Werner}). Moreover, it is evident how many states may
cross the boundary between entangled and separable states. (b)
Distribution of $e_m([\hat{\rho}_4^i]^{(\tau)})$ for $10^3$ simulations
of the target state $\overline{\hat{\rho}_4}$ as a function of the
fidelity $F(\hat{\rho}_4^i,\hat{\rho}_4^{(w)})$ (green dots). The same
distribution as a function of
$F([\hat{\rho}_4^i]^{(w)},\hat{\rho}_4^{(w)})$ (orange dots), matches
with the theoretical parametric curve (dashed black curve) obtained by
evaluating $F$ and $e_m$ of the approximated Werner states. The values
of entanglement for the average state $\overline{\hat{\rho}_4}$ and for
the target Werner state $\hat{\rho}_4^{(w)}$, are compatible (black dots
and error bars).} \label{f:negativity}
\end{figure}
\subsubsection{Entanglement}
The separability of two-qubit systems is established by the
Peres-Horodecki criterion \cite{per,hor}: a quantum
state of two qubits $\hat{\rho}$ is separable if and only the partially
transposed density matrix is positive, i.e. $\hat{\rho}^{(\tau)}\geq 0$.
Thus, it is possible to study entanglement or separability properties by
evaluating the eigenvalues of the partially transposed density matrix.
We compute the minimum of these eigenvalues \begin{equation}
e_m(\hat{\rho}^\tau)\equiv \text{min} \{\lambda_n^\tau\}_{n=1}^4,
\end{equation}
for both the considered strategies, i.e. for the distributions of
resampled states $\hat{\rho}_k^i$ and of the approximated Werner states
$[\hat{\rho}_k^i]^{(w)}$.  For a Werner state the minimum
eigenvalue, which may assume negative values, is given by
$e_m([\hat{\rho}^{(w)}]^\tau)=(1-3p)/4$. 
\par
In Fig.~\ref{f:negativity}~(a) we plot the distribution of
$e_m(\hat{\rho}^\tau)$ as a function of the Werner parameter $p$ for all
the $N_{\text{MC}}=10^3$ resampled states, with average tomographic
states $\overline{\hat{\rho}_k}$. We note that the average states
arrange along the curve of the theoretical behavior for a Werner state
and that the resampled states follow the same prediction (see
Table~\ref{t:Werner}). Nonetheless, there is evidence of some 
states generated from a separable experimental state in the entangled
region, and viceversa. This first observation reveals that statistical
fluctuations in an experiment may still produce quantum states with
properties radically different from the expected ones, such as
separability and entanglement.  \begin{figure}
\subfigure[]{\includegraphics[width=0.4\textwidth]{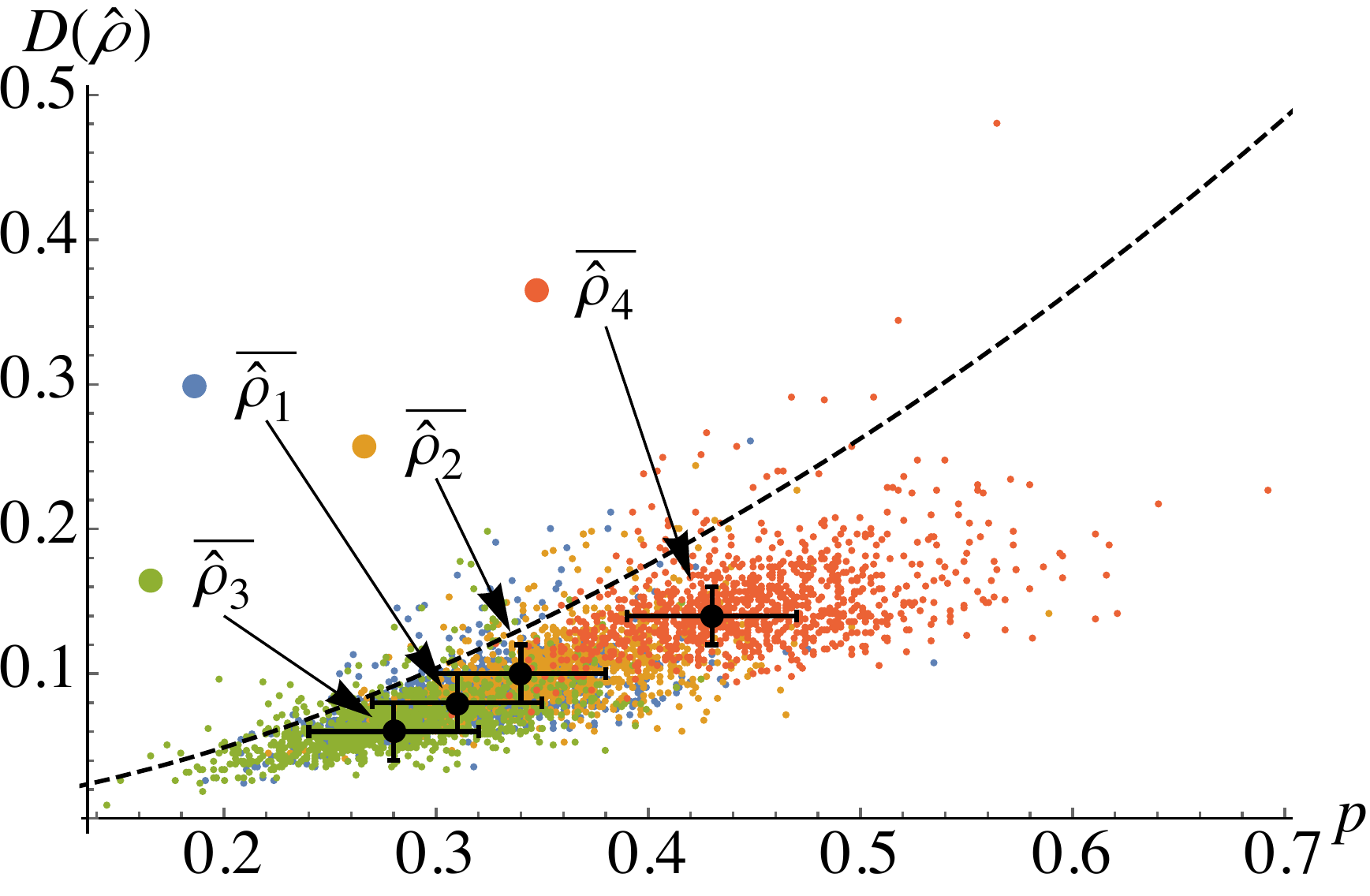}} 
\subfigure[]{\includegraphics[width=0.4\textwidth]{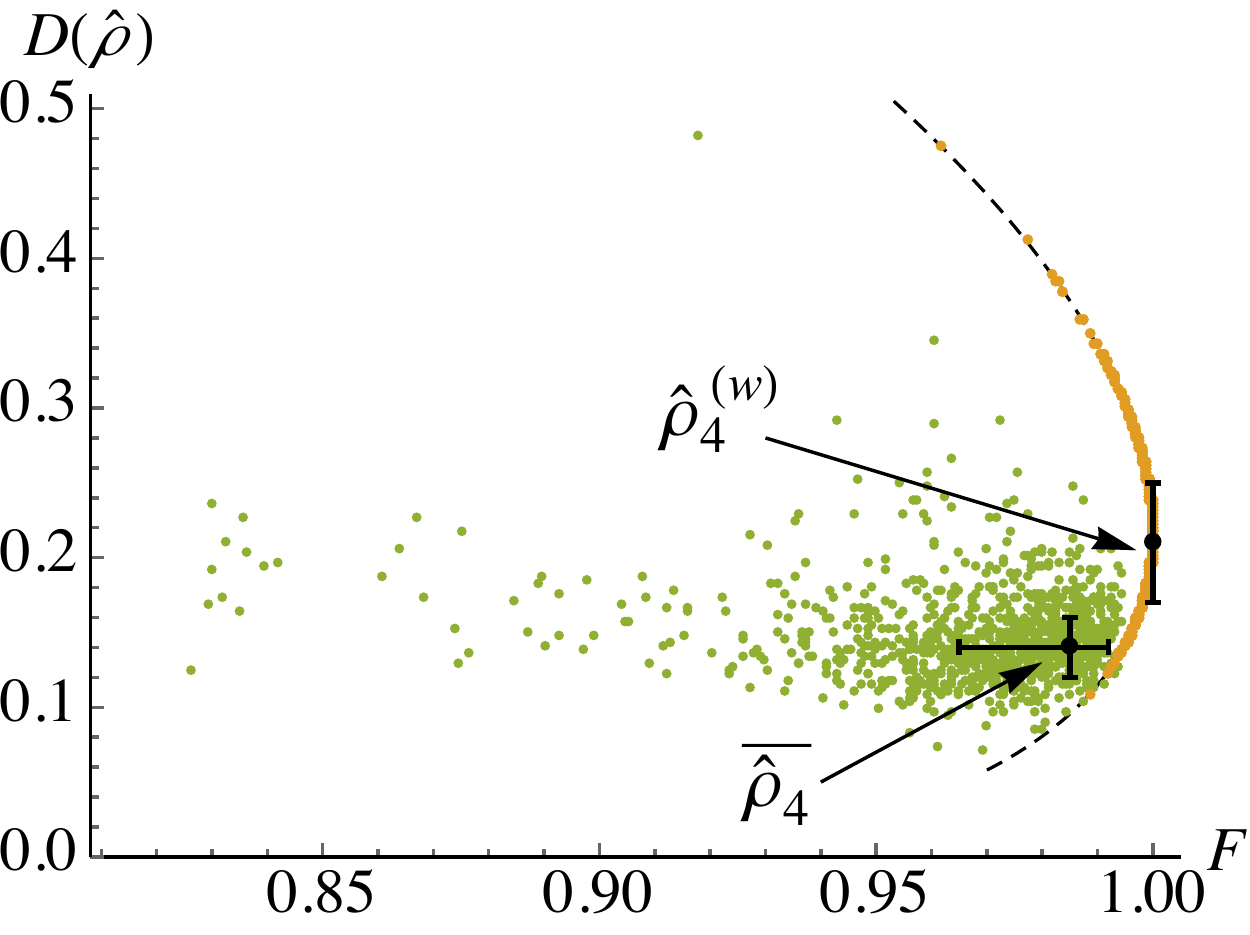}} 
\caption{(Color online)  (a) Distribution of $D(\hat{\rho}_k^i)$ for
$10^3$ resampled states as a function of the Werner parameter $p$. The
$N_{\text{exp}}=4$ average states $\overline{\hat{\rho}_k}$ are
highlighted with black dots and error bars. The theoretical curve
$D(\hat{\rho}^{(w)})$ (dashed black line), relative to the discord of
the ideal Werner state (\ref{Werner}), is systematically higher than the
discord distribution of the tomographic states $\hat{\rho}_k^i$.  (b)
Distribution of $D(\hat{\rho}_4^i)$ for $10^3$ simulations of the target
state $\overline{\hat{\rho}_4}$ as a function of the fidelity
$F(\hat{\rho}_4^i,\hat{\rho}_4^{(w)})$ (green dots). The same
distribution as a function of
$F([\hat{\rho}_4^i]^{(w)},\hat{\rho}_4^{(w)})$ (orange dots), matches
with the theoretical parametric curve (dashed black curve) obtained by
evaluating $F$ and $D$ of the approximated Werner states. The values of
quantum discord for the average state $\overline{\hat{\rho}_4}$ and for
the target Werner state $\hat{\rho}_4^{(w)}$, are not compatible (black
dots and error bars).}
\label{f:discord}
\end{figure}
In Fig.~\ref{f:negativity}~(b), we focus on the most entangled state
$\overline{\hat{\rho}_4}$ and compare the distribution of the resampled 
states and of the corresponding Werner states, in terms of
$e_m(\hat{\rho}^\tau)$ and fidelity with the target state
$\hat{\rho}_4^{(w)}$. In this way we can highlight the differences
between the two possible strategies for data analysis, i.e. between the
evaluation of the properties of the direct tomographic states and the
approximated Werner states. This second strategy compel the resampled
states to follow the single-parameter Werner state (\ref{Werner}) and,
thus, to force the distribution of the least eigenvalue
$e_m([\hat{\rho}_k^{(w)}]^{(\tau)})$ according to the theoretical
prediction (dashed black curve in the plot). In this way, we obtain a
distribution of Werner states having, obviously, very high values of
fidelity with the target (Werner) state. On the other hand, tomographic
states display lower values of fidelity with the target state, but
$e_m(\hat{\rho}^\tau)$ is evaluated directly from the tomographic
density matrices. From the statistical analysis, we conclude that the
estimated value of $e_m(\hat{\rho}^\tau)$ is compatible within errors
for both the adopted strategies. We will see in the following that these
two strategies may lead to different and non-compatible results for
another property of quantum states, the quantum discord.

\subsubsection{Quantum discord}
Another widely adopted measure of the amount of quantum correlations in
a state $\hat{\rho}$ is the quantum discord \cite{disc:1,disc:2}, which is defined
starting from two equivalent definitions, in the classical domain, of the mutual
information, i.e. the total amount of correlations of $\hat{\rho}$:
\begin{subequations}\label{MI}\begin{align}
\mathcal{I}(\hat{\rho})&=S(\hat{\rho}_A)+S(\hat{\rho}_B)-S(\hat{\rho}),\label{MI1}\\
\mathcal{J}_A(\hat{\rho})&=S(\hat{\rho}_B)-\min\sum_j c_j S(\hat{\rho}_{B|j}),\label{MI2}
\end{align}\end{subequations}
where $\hat{\rho}_A$ ($\hat{\rho}_B$) is the reduced density matrix of
$\hat{\rho}$ for the subsystems $A$ (or $B$) and $S(\hat{\rho})=-\text{Tr}[\hat{\rho}\log_2(\hat{\rho})]$
the von Neumann entropy.
While the first definition (\ref{MI1}) is based only on the von Neumann entropy,
the second one in Eq.~(\ref{MI2})
accounts for all the classical correlations that can be
detected by local projective measurements only on a subsystem. Here
$\hat{\rho}_{B|j}=\text{Tr}_{\text{A}}[\hat{\Pi}_j\hat{\rho}\,\hat{\Pi}_j]/\pi_j$
is the reduced state of $B$ conditioned to the set of projectors
$\{\hat{\Pi}_j\}$, with probability
$\pi_j=\text{Tr}[\hat{\Pi}_j\hat{\rho}\,\hat{\Pi}_j]$ for the outcome
$j$, and the minimum in Eq.~(\ref{MI2}) is taken over all the possible
$\{\hat{\Pi}_j\}$. A similar definition applies for local measurements on
subsystem $B$. The quantum discord is then evaluated as the residual
information stemming from the difference of the two definitions in Eqs.
(\ref{MI}), which has a pure quantum character:
\begin{equation}\label{discord}
D(\hat{\rho})\equiv \mathcal{I}(\hat{\rho}) - \mathcal{J}_A(\hat{\rho}).
\end{equation}
For a Werner state (\ref{Werner}) the quantum discord can be analytically evaluated:
\begin{equation}\label{discordW}\begin{split}
D(\hat{\rho}^{(w)})=&\frac{1+3p}{4}\log_2(1+3p)+\frac{1-p}{4}\log_2(1-p)\\
&-\frac{1+p}{2}\log_2(1+p).
\end{split}\end{equation}
\begin{figure}[t]
\includegraphics[width=0.9\columnwidth]{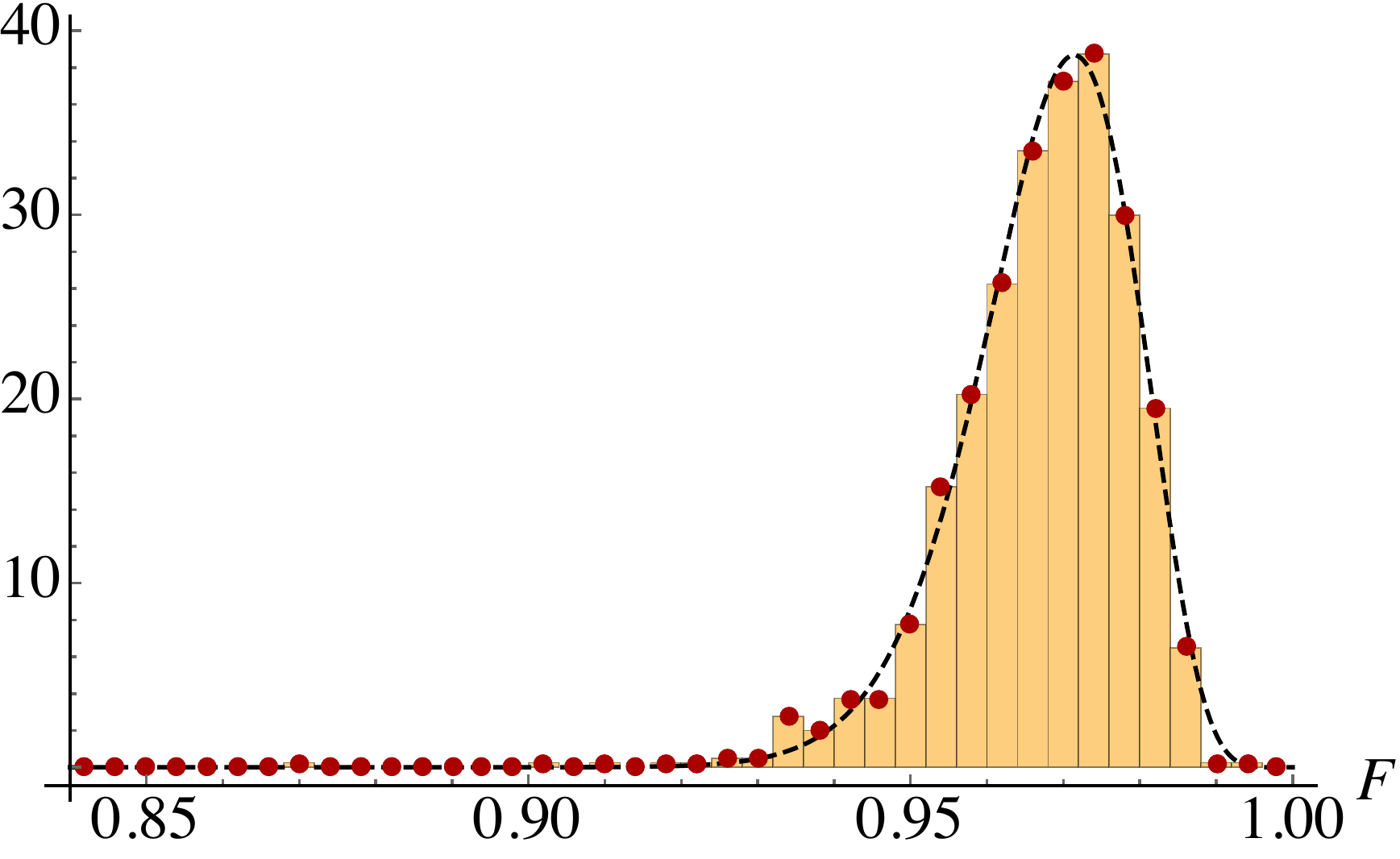}
\caption{The histogram shows the distribution of
$F(\hat{\rho}_3^i,\hat{\rho}_4^{(w)})$, i.e. the fidelity between a set
of separable states and an entangled target state. The data are well
fitted by a $\beta$-distribution, centered around $F\approx 0.97$.}
\label{f:histogram}
\end{figure}
\par
In Fig.~\ref{f:discord}(a) we plot the quantum discord
$D(\hat{\rho}_k^i)$ for all the $N_{\text{MC}}=10^3$ resampled states,
with average tomographic states $\overline{\hat{\rho}_k}$, as a function
of the Werner parameter $p$. The theoretical behavior of
Eq.~(\ref{discordW}), represented in the figure by a dashed curve, puts
in evidence the high discrepancy with the quantum discord computed for
the tomographic states. The approximation to Werner states  leads to a
systematic over-estimation of the quantum discord for the
two-qubit states (see also Table~\ref{t:Werner}). We can enforce this
result by looking at the distribution of the quantum discord relative to
the single set of states $\hat{\rho}_4^i$ only, as a function
of the fidelity with the corresponding target Werner state
$\hat{\rho}_4^{(w)}$ (see Fig.~\ref{f:discord}(b)). Most of the
states are contained in a region of high values of fidelity, thus
suggesting that the approximation to an average Werner state could be
correct. Nonetheless, if we consider the distribution of the
approximated Werner states, we observe that the values of fidelity
increase, but the value of quantum discord of the target Werner state
$\hat{\rho}_4^{(w)}$ is out of the limits of compatibility with the
average quantum discord $D(\overline{\hat{\rho}_4})$. This suggest that
the second strategy employing the approximation to Werner states reveals
to be too drastic, as it does not account properly for the actual
tomographic reconstruction of the density matrix.  \par
We can conclude that, even though high values of fidelity between a
target state and a tomographic state are achieved, the properties of the
two can be very different. On an extreme level, we can look at the
distribution of the fidelity between the most classical states we
generated, $\hat{\rho}_3^i$, and, at the opposite, the most entangled
one as the target state, namely $\hat{\rho}_4^{(w)}$. The probability
density histogram in Fig.~\ref{f:histogram} shows that the two kind of
states should result compatible with a level of fidelity $F\approx
0.97$, even though they possess clearly different properties.
\section{Conclusions} \label{s:conclusions}
In conclusion, we have addressed quantum state reconstruction for DV
and CV quantum optical systems and
experimentally analyzed the significance of fidelity as a figure of
merit to assess the properties of the reconstructed state.  State
reconstruction, in the two cases, has been performed adopting homodyne
and MLE tomography techniques. One of the most natural ways to link the
tomographic results to the target states, i.e. the quantum states
supposed to be generated by the designed experimental setup, is the
evaluation of fidelity. In order to study the relation between fidelity
and the experimental states, we performed statistical analysis using
Monte Carlo sampling of each experiment, re-generating sets of
$N_{\text{MC}}=10^3$ data samples and analyzing the distribution of
some of their main properties as a function of fidelity.  
\par 
In the CV framework, we employed a thermal-state
seeded OPO cavity, an experimental configuration which allows to
generate STS on-demand. The accurate control of the thermal and
squeezing component of the apparatus, allows us to address
the quantum-to-classical transition for these states. 
Our results show that even for high values of fidelity and imposing energy
constraints, one may find neighboring states in terms of fidelity
which, however, not share the same quantum/classical properties.  
\par 
In the DV context, we experimentally obtained pairs of
polarized photons from type-I downconversion and conveniently generate
the Werner mixed states. This one-parameter family of states allowed us
to analyze the non-classical properties of two-qubit states in terms of
entanglement/separability and to evaluate the amount of quantum
correlations by means of the quantum discord. We found that 
a fidelity based approximation of the tomographic states by Werner
states may lead to an overestimation e.g. of the quantum
discord. Moreover, high values of fidelity may occur between two very
different states in terms of their separability properties.  
\par 
Overall, we conclude that while fidelity is a good measure of
geometrical proximity in the Hilbert space it should not be 
used as the sole benchmark to certify quantum properties, 
which should be rather estimated tomographycally in a direct way.

\section*{Acknowledgments}
This work has been supported by UniMI through the
UNIMI14 grant 15-6-3008000-609 and the H2020
Transition Grant 15-6-3008000-625, and by EU through the H2020
Project QuProCS (Grant Agreement 641277).
\appendix

\section{Evaluation of uncertainties by Monte Carlo resampling}
\label{s:MC}
In order to avoid the limitations of finite samples and the 
influence of systematic unpredictable errors that could be 
present in an experiment, we evaluate uncertainties by Monte
Carlo resampling of data, according to standard metrological 
prescriptions \cite{JCGM} valid for any statistical models 
having a single output quantity and input quantities with 
arbitrary distribution. Here we provide a brief summary of 
the main assumptions and principles.
\par
The measured quantities of interest $X_i$ are random variables 
distributed according to a given probability density function 
(PDF) $\mathcal{G}(X_i)$. In particular, we assume normal 
distributions characterized by mean value $\langle x_i \rangle$ 
and standard deviation $\delta x{_i}$. Monte Carlo evaluation of 
uncertainties is based on sampling random outcomes from 
$\mathcal{G}(X_i)$ according to experimental data, which themselves 
fix the the average values and the standard deviations. 
In particular, as described in the main text, the considered 
experimental measurements, for CV systems, correspond to 
homodyne detection of the radiation field quadratures, whereas 
for DV systems we perform coincidence photon counting 
measurements of polarized photons. Starting from experimental 
results, we generate $N_{\text{MC}}=10^3$ resampled replicas of the 
experiments, thus building a significative sample for the 
statistical analysis.

\end{document}